%% file: paper.tex
\documentclass[12pt,preprint]{aastex}

\def\lea{\mathrel{<\kern-1.0em\lower0.9ex\hbox{$\sim$}}}
\def\gea{\mathrel{>\kern-1.0em\lower0.9ex\hbox{$\sim$}}}

\slugcomment{}

\shorttitle{M31 RR Lyraes}
\shortauthors{Sarajedini et al.}

\begin{document}


\title{RR Lyrae Variables in Two Fields in the Spheroid of M31\footnote{Based on observations
taken with the NASA/ESA Hubble Space Telescope, obtained at the 
Space Telescope Science Telescope.} }

\author{Ata Sarajedini and Conor L. Mancone}
\affil{Department of Astronomy, University of Florida, Gainesville, FL 32611}
\email{ata@astro.ufl.edu, cmancone@astro.ufl.edu}

\author{Tod R. Lauer}
\affil{National Optical Astronomy Observatory\footnote{The National Optical
Astronomy Observatory is operated by AURA, Inc., under cooperative agreement
with the National Science Foundation.},
P.O. Box 26732, Tucson, AZ 85726}
\email{tlauer@noao.edu}

\author{Alan Dressler and Wendy Freedman}
\affil{Observatories of the Carnegie Institution of Washington, Pasadena, CA 91101}
\email{dressler@ociw.edu, wendy@ociw.edu}

\author{S. C. Trager}
\affil{Kapteyn Astronomical Institute, University of Groningen, NL-9700 AV
Groningen, Netherlands}
\email{sctrager@astro.rug.nl}

\author{Carl Grillmair}
\affil{Spitzer Science Center, Pasadena, CA 91125}
\email{carl@ipac.caltech.edu}

\author{Kenneth J. Mighell}
\affil{National Optical Astronomy Observatory$^2$,
P.O. Box 26732, Tucson, AZ 85726}
\email{kmighell@noao.edu}



\begin{abstract}
We present Hubble Space Telescope observations taken with the Advanced
Camera for Surveys Wide Field Channel of two fields near M32 - between 
four and six kpc from the center of M31. The data cover a time baseline sufficient
for the identification and characterization of 681 RR Lyrae variables of
which 555 are ab-type and 126 are c-type. The mean magnitude of these stars is 
$\langle$$V$$\rangle$$=25.29 \pm 0.05$ where the uncertainty combines both
the random and systematic errors. The location of the stars in the Bailey Diagram
and the ratio of c-type RR Lyraes to all types are both closer to
RR Lyraes in Oosterhoff type I globular clusters in the Milky Way as compared with
Oosterhoff II clusters. The mean periods of the ab-type and c-type RR Lyraes
are $\langle$$P_{ab}$$\rangle$$=0.557\pm0.003$ and
$\langle$$P_{c}$$\rangle$$=0.327\pm0.003$, respectively, where the uncertainties 
in each case represent the standard error of the mean.
When the periods and amplitudes of the ab-type RR Lyraes in our
sample are interpreted in terms of metallicity, we find the metallicity
distribution function to be indistinguishable from a Gaussian with a 
peak at $\langle$[Fe/H]$\rangle$=$-1.50\pm0.02$, where the quoted uncertainty is 
the standard error of the mean.
Using a relation between RR Lyrae luminosity and metallicity
along with a reddening of $E(B-V) = 0.08 \pm 0.03$, 
we find a distance modulus of $(m-M)_0=24.46\pm0.11$ for M31. 
We examine the radial metallicity gradient in the
environs of M31 using published values for the bulge and halo of M31
as well as the abundances of its dwarf spheroidal companions and globular
clusters. In this context, we conclude that the RR Lyraes in
our two fields are more likely to be halo objects rather than associated
with the bulge or disk of M31, in spite of the fact that they are located
at 4-6 kpc in projected distance from the center.
\end{abstract}



\keywords{stars: variables: other -- galaxies:  stellar content -- 
galaxies: spiral -- galaxies: individual (M31) }

\section{Introduction}

Pulsating variable stars such as RR Lyraes are powerful probes useful
for investigating the properties of stellar populations. The mere presence of RR Lyraes
among a population of stars suggests an ancient origin since ages older than 
$\sim$10 Gyr are required to produce RR Lyrae variables. Their periods and
amplitudes are a reflection of the metal abundance of the population. Along
with their incredible usefulness, RR Lyraes are also relatively straightforward to
identify and characterize. This is because of their short periods and the distinct light curve
shapes of the ab-types, which pulsate in the fundamental mode and 
exhibit a relatively rapid rise to maximum
and a gradual decline to minimum. This is in contrast to the c-type RR Lyraes 
which pulsate in the first harmonic and show light curves that are more akin to sine curves.
In spite of the great potential RR Lyraes hold as astrophysical tools, they have
not been widely studied in our nearest large neighbor galaxy, Andromeda.

One of the first studies attempting to identify RR Lyraes in M31 was that of Pritchet
\& van den Bergh (1987). They used the Canada-France-Hawaii 3.6m telescope
to observe a field at a distance of 9 kpc from the center of M31 along the minor
axis partially overlapping the field observed by Mould \& Kristian (1986). 
They identified 30 RR Lyrae candidates and were able to estimate periods for 28 of them.
These ab-type variables have a mean period of 
$\langle$$P_{ab}$$\rangle$=0.548 days. The
photometric errors in their data prevented them from identifying the lower-amplitude
c-type RR Lyraes. 

The RR Lyrae variables in M31 globular clusters have been investigated by
Clementini et al. (2001, {\it cf.} Contreras et al. 2008). They used the 
Wide Field Planetary Camera 2 (WFPC2)
onboard the Hubble Space Telescope (HST) to make the first tentative
detection of RR Lyraes in G11, G33, G64, and G322, finding  two, four, 
11, and eight variables, respectively. Detection and characterization of 
these stars in globular clusters is more challenging than in the M31 field
because of the increased crowding.

Returning to the work on field RR Lyraes, 
Dolphin et al. (2004) observed the same field as Pritchet
\& van den Bergh (1987) using the WIYN 3.5m on Kitt Peak. They found
24 RR Lyrae stars with a completeness fraction of 24\%, suggesting that their
$\sim$100 square arcmin field could contain $\sim$100 RR Lyraes resulting in a
density of one RR Lyrae per square arcmin. This is much less than the value of
$\sim$17 per square arcmin found by Pritchet \& van den Bergh (1987). They also
noted for the first time that the mean metallicity of the M31 RR Lyraes seemed to
be significantly lower than that of the M31 halo. The work of Durrell et al.
(2001) had reported a peak value of $[M/H]$$\sim$--0.8 for the M31 halo.
Dolphin et al. (2004) were not able to reconcile this abundance value with the distance
implied by the mean magnitude of their RR Lyrae sample. 

The first definitive work on the RR Lyraes of M31 was published by Brown et al.
(2004, hereafter B2004) and made use of $\sim$84 hours of imaging time 
(250 exposures over 41 days) with
the Advanced Camera for Surveys onboard HST. Their 
field was located along the minor axis of M31 approximately 11 kpc from
its center. Their analysis revealed a complete sample of RR Lyrae stars
consisting of 29 ab-type variables and 25 c-type. The periods of these
stars suggest a mean metallicity of $[Fe/H]$$\sim$-1.6 for the old
population in the Andromeda halo. This is qualitatively consistent with
the assertions of Dolphin et al. (2004) regarding the metal abundance of
the M31 halo - that it is lower than the value suggested by the work of
Durrell et al. (2001). More recent work has shown that the M31 halo extends
from 30 to 165 kpc (Guhathakurta et al. 2005; Irwin et al. 2005) and has a
metallicity that is actually closer to that of the Milky Way halo (Kalirai et al. 2006;
Koch et al. 2008). 

There is one more paper of note related to this topic and that is the work of 
Alonso-Garc\'{i}a et al. (2004). They used the Wide Field Planetary Camera 2
onboard HST to image a field $\sim$3.5 arcmin to the east of M32 and compared
it with a control field that samples the M31 field stars well away from M32. 
They identify variable stars that they claim are RR Lyraes belonging to 
M32 therefore suggesting that M32 possesses a population that
is older than $\sim$10 Gyr. They were not able to classify the RR Lyraes or
derive periods and amplitudes for them so their results are not directly comparable to
ours.

This review of the literature reveals a significant deficiency in the spatial
coverage of RR Lyrae studies
in the vicinity of M31. Given the great astrophysical
utility of RR Lyrae variables and the expansive size
of M31 on the sky, it is clear that a survey of these stars sampling a diversity of
regions in Andromeda will provide valuable insights into the star formation and
chemical enrichment history of our nearest spiral neighbor.

With this in mind, this paper presents 
the results of archival HST/ACS observations showcasing
the RR Lyrae population in the inner regions of the M31 spheroid. The next section
describes the observational material and the photometric procedure. We move
on to describe the technique used to identify and characterize the variable stars
in Sec. 3. The results of this study are described in Sec. 4, and a discussion of
these results within the broader context of the M31 halo are presented in Sec. 5.
Our conclusions are then summarized in Sec. 6.

\section{Observations}

The observations used in the present study were obtained with the Hubble Space
Telescope Advanced Camera for Surveys (HST/ACS) in parallel with the imagery
conducted by program GO-10572.   The primary goal of this program was to obtain a
deep color-magnitude diagram of the envelope of M32, using the High Resolution
Channel (HRC) of the ACS.  The total envelope exposure was 32 orbits, split between
two filters.
An identical set of HRC exposures was later obtained in a background field selected
to represent the M31 disk+halo contribution to the M32 envelope exposures.  The
present images are the parallel observations associated with each pointing obtained
with the ACS Wide Field Channel (WFC) using the F606W ($\sim$V) filter.  Table 1
provide some details of the observational data. The temporal coverage is 2.2 days
for field 1 and 3.1 days for field 2.

Figure 1 shows the locations of these fields relative to M31.   While the
M32-background field was carefully selected to fall along the M31 isophote that ran
through the M32 envelope field, the two fields were significantly separated in time,
thus different spacecraft rolls between the two epochs caused the parallel WFC
aperture to fall randomly about the primary HRC fields. Field 1 thus samples a
region that is 4.5 kpc in projected distance from the center of M31, while 
Field 2 is located 6.6 kpc from
the center.  Note that throughout this paper, we adopt an M31 distance modulus of
$(m-M)_o = 24.43$ corresponding to a distance of 770 kpc (Freedman \& Madore
1990).

The spacecraft was dithered in a complex pattern to both achieve Nyquist-sampling in
the HRC and rejection of CCD defects.  Each pair of orbits was dithered in a
$2\times2$ square pattern of 0.5 HRC pixel steps, followed by larger steps to trace
a skewed-square spiral of $\sim0.2$ arcsec total amplitude over the 16 total orbits
devoted to each filter/field combination.  The subpixel dithering to achieve Nyquist
sampling in the HRC is not optimal for the $2\times$ larger pixels of the WFC, but
the larger-scale dither pattern fortunately served to offer diverse sampling
information for the parallel imagery.


\section{Reductions}

\subsection{Photometry of Program Frames}

We chose to work on the FLT images as retrieved from the HST archive. These
frames have been bias-subtracted and flat-fielded, but, unlike the drizzled (DRZ)
images, they retain the geometric distortions of ACS
(Mahmud \& Anderson 2008). Photometry was performed
using the same procedure as Sarajedini et al. (2006). The first step is the application
of geometric correction images to the Wide Field Channel 1 and 2 (WFC1 and WFC2,
respectively) portions of the FLT images. 
After this step, the data quality maps are applied where the values of the bad
pixels in the science images are set to a number well below the sky background 
to be sure the photometry software ignores those pixels. At this point, the resultant
images are ready to be photometered.

The detection of the stellar profiles and the measurement of magnitudes was
done with the DAOPHOT/ALLSTAR/ALLFRAME crowded-field photometry 
software (Stetson 1987; 1994). After the application of the standard FIND and
PHOT routines to detect stars and perform aperture photometry on them, 
ALLSTAR was applied to each of the 32 images in order to derive 
well-determined positions for all of the stars. In this step and in subsequent ones
involving the application of a point-spread function (PSF) in order to determine
positions and magnitudes, we made use of the high signal-to-noise PSFs
constructed by Sarajedini et al. (2006). The reader is referred to that paper
for the details of the PSF construction process. 

The stellar positions from the ALLSTAR runs were used to construct a coordinate
transformation between each of the 32 images and these were used to combine
all of the images into one master frame per field. This combined frame was then input into
ALLSTAR, from which a master coordinate list of stellar profiles was constructed.
The resultant coordinate list along with the spatial transformation between the
images and the PSFs were used in ALLFRAME to derive magnitudes for
all detected profiles  on each image. At this point, the measurements on each
of the individual frames were matched and only stars appearing on all 32 images
were kept. 

The standardization of the individual magnitudes proceeded in the following
manner. First, the correction for the charge transfer efficiency problem was
applied using the prescription of Reiss \& Mack (2004). The magnitudes were
then adjusted to a radius of 0.5 arcsec and corrected for exposure time.
Offsets to an infinite radius aperture published by Sirianni et al. (2005)
were then applied. Finally, the resultant values
were calibrated to the VegaMAG system using the zeropoint for the F606W
filter from Sirianni et al. (2005). A correction to this zeropoint amounting to
0.022 mag was applied as a result of 
a revised calibration of the ACS/WFC photometric performance by Mack
et al. (2007).
Each of our magnitude measurements is affected by three sources of
systematic error:
the uncertainty in the aperture corrections ($\pm$0.02 mag), the error in
the correction to infinite aperture ($\pm$0.00 mag) for the F606W filter,
and the error in the VegaMAG zeropoint ($\pm$0.02 mag).

\subsection{Characterization of the Variable Stars}

For a given star with 32 magnitudes measurements, we calculated the mean
photometric error as returned by ALLFRAME 
($\langle$$\sigma_{err}$$\rangle$) and the standard deviation of the
measurements ($\sigma_{sd}$). For the first round of variable searching,
we considered any star a candidate if 
$\sigma_{sd}$~/~$\langle$$\sigma_{err}$$\rangle$$\geq$3.0. Approximately 
3000 stars fit this criterion in each of our two fields. 

These stars were then input into our template fitting period-finding algorithm, which
is based on the method used by Sarajedini et al. (2006) as originally formulated by
Layden \& Sarajedini (2000). We have taken the FORTRAN
code written for the Layden \& Sarajedini (2000) study and rewritten
it using the Interactive Data Language (IDL) incorporating a graphical user 
interface (GUI). The original FORTRAN code used the `amoeba' minimum-finding
algorithm exclusively, but our code, dubbed
FITLC\footnote{ http://www.astro.ufl.edu/$\sim$cmancone/fitlc.html}, 
has the option to use a more robust algorithm known as `pikaia' which has its
roots in the study of genetics. The software uses 10 template light curves -
six ab-type RR Lyraes, two c-type RR Lyraes,  one eclipsing
binary, and one contact binary. It searches over a period range from 0.2 day to
a specified maximum (2.2 days for our field 1 and 3.1 days for field 2)
looking for the period that minimizes the value of $\chi^2$.
This is accomplished with
a two step process.  First pikaia is used to find the combination of
epoch, amplitude, and mean magnitude that minimize $\chi^2$ at evenly spaced
period increments of 0.01 day.  Then pikaia is applied again to find the
combination of epoch, amplitude, mean magnitude, and period that minimize
$\chi^2$ within $\pm$ 0.01 day of the period with the lowest $\chi^2$.  The best
fitting period from this final step is taken to be the period of the variable.
The resultant phased light curves for each star were 
visually examined, and the stars that presented a compelling case for
variability were retained in our final database. Of the $\sim$6000 total stars 
originally fit, 752 exhibit genuine variability as shown by our data.

As a test of our template-fitting method, we have also applied the Lomb-Scargle
period-finding algorithm (Scargle 1982; Horne \& Baliunas 1986) 
to the time series photometry of the RR Lyraes in our sample. We find a
mean difference of 0.0007d in the periods determined by the two methods
throughout the period range of RR Lyraes. In the minority of cases where
template-fitting and Lomb-Scargle yield significantly different results, the
resultant phased light curves are of significantly higher quality for the
former method as compared with the latter. Furthermore, to test for the 
presence of aliasing effects in our derived periods, we have also examined
fitted light curves using periods that correspond to $\chi^2$ minima near
half of the optimum period. In all cases, these
fits are clearly inferior to the ones yielded by the optimum period
from FITLC.

Tables 2 and 3 list the individual F606W magnitudes of each variable at 
each epoch wherein 2 450 000 has been subtracted from the epoch value
while Tables 4 and 5 list the candidate variables in our two fields
along with their properties such as period, amplitude, and mean intensity-weighted
magnitude. These stars fall into two broad categories. First, there are 
those that clearly show variability, 
but their periods are comparable to or longer than our observing window.
These are referred to as `long period' in Tables 4 and 5. There is also one candidate
anomalous cepheid in our dataset, which is so indicated in Table 5.
The second category includes stars  that exhibit clear periodic
variability with a period that is significantly shorter than our observing window
but longer than 0.2d. These are
the stars for which we can be confident of our periods. 
Some examples of stars with periods longer than
our observing window are shown in Fig. 2 while Fig. 3 shows phased light curves of 
a number of contact and eclipsing binaries in our dataset. Figure 4 displays the
phased light curves of all of the RR Lyrae variables for which we have derived periods.
\footnote{The complete figure is only available in the electronic version of the journal.}

All of the stars that exhibit RR Lyrae light curves also have apparent magnitudes in the 
range one would expect if they are at the distance of M31. Therefore, it is reasonable
to assume that most if not all of these objects are RR Lyrae stars belonging to M31 
and/or its environs. This assertion will become clearer when we compare the luminosity 
function (LF) of the non-variable stars with that of the RR Lyraes. 

It should be noted that we are much less confident about the properties
of the eclipsing and contact binaries that we have identified as compared
with the RR Lyraes. This is because we know what period range to expect for
the RR Lyraes ($\sim$0.25d to $\sim$0.90d), so that our observing window
provides coverage of multiple cycles of variation for a given RR Lyrae star.
In contrast, the periods of the eclipsing and contact binaries cannot be similarly
constrained so it is difficult for us to ensure that our observing window is
sufficient to derive the periods of these variables. As such, we have provided
information for these stars (positions, periods, amplitudes, and magnitudes) so that
future observers can confirm the nature of their variability, but we will not
consider them further here. Instead, for the
remainder of this paper, we will limit our discussion to the 681 RR Lyrae variables
in our sample and what they reveal about the properties of the M31 system.

\subsection{Light Curve Simulations}

In order to characterize the possible biases in the derived
periods of our variable star sample, we have performed simulations of our
light curve fitting technique in the following manner.
For the ab-type RR Lyraes, we selected 
one of the light curve templates 
(the results are insensitive to the actual ab-type template used) and 
produced artificial variables with a period range of 0.45 to 0.80 days and amplitudes 
between 0.3 and 1.3 mag. For the c-type RR Lyraes, a period range of 0.25 to 
0.40 days and an amplitude range of 0.2 to 0.5 mag was used. 
One thousand 
variables were generated in each case and the mean photometric error
at the level of the RR Lyraes was used to populate the light curves 
using the same observing window as the actual data. These simulated
light curves were input into our template light curve fitting software.
We are interested in comparing the 
input periods with the output periods in order to gauge any possible biases
present in our analysis method.


Figures 5 (RRab) and 6 (RRc) show the result of
these fits for Field 1 wherein the upper panel shows the mean 
difference between the input and output periods while the lower panel
 illustrates the differences in the period distributions. The result of the 
 simulations for RR Lyraes in Field 2 are indistinguishable from those in Field 1. 
Of the 1000 ab-type RR Lyraes generated, none were mistaken for any other
type of variable among the 10 light curve templates used in the fitting. For
the c-type light curves, 2 of the 1000 generated variables were fit with
a contact binary template. We find no significant biases in our
determination of the periods for both types of RR Lyraes. As such, we will
not apply any sort of correction to our derived periods. As for the errors in the period
and amplitude determinations, these simulations suggest that an individual ab-type 
RR Lyrae has an error of $\pm$0.005 day and $\pm$0.044 mag, respectively. 
For a c-type RR Lyrae star, these error values are $\pm$0.001 day and 
$\pm$0.022 mag.

\section{Results}

\subsection{Luminosity Functions}

Figure 7 displays a comparison of the LFs of the nonvariable stars in the two
fields. Both distributions feature a quick rise from brighter magnitudes to
fainter ones with a pronounced peak at $m_{F606W}$$\sim$25.3 representing
the core-helium burning horizontal branch (HB) stars. A sudden
drop in both LFs at $m_{F606W}$$\sim$27.7 suggests the onset of significant
incompleteness as the limit of the photometry is approached. The fact that
the HB stars are more than 2 magnitudes brighter than this completeness
threshold indicates that our sample of RR Lyrae variable candidates should 
not be adversely biased by photometric incompleteness.

The two panels of Fig. 8 compare the LF of the nonvariable stars with those
of the RR Lyraes in the two observed fields. We have used the intensity-weighted
magnitudes listed in Tables 4 and 5 to construct these distributions.
Gaussian fits to the regions around the RR Lyrae LF peaks yield 
$\langle$$m_{F606W}$$\rangle$$=25.20 \pm 0.04$ for Field 1 and
$\langle$$m_{F606W}$$\rangle$$=25.22 \pm 0.04$ for Field 2, where the errors
represent the standard error of the mean combined with the uncertainty in the
photometric zeropoint added in quadrature. For the nonvariable stars, these
peaks correspond to 
$\langle$$m_{F606W}$$\rangle$$=25.25 \pm 0.04$ for Field 1 and
$\langle$$m_{F606W}$$\rangle$$=25.23 \pm 0.04$ for Field 2. The 1-$\sigma$
width of these distributions is $\sim$0.11 mag. These numbers
suggest no significant difference in the mean magnitudes of the RR Lyraes
and the nonvariable stars. This serves to confirm our assertion that most if not all of
the variable stars in this magnitude range are RR Lyrae variables. In addition,
when we combine the RR Lyrae variables from both fields, we find a mean
magnitude of $m_{F606W}$$ = 25.21 \pm 0.01$ on the VEGAmag system. 
The quoted uncertainty represents the standard error of the mean. When 
converted to the 
V-band using the mean offset for RR Lyraes in the middle of the instability
strip from B2004 of $V-m_{F606W}$ = 0.08 $\pm$ 0.04, we derive 
$\langle$$V(RR)$$\rangle$$=25.29 \pm 0.01$ (random) $\pm 0.05$ (systematic). 
This compares favorably
with the value of $\langle$$V(RR)$$\rangle$$=25.30 \pm 0.01$ based on the average
for the ab-type and c-type RR Lyraes from the B2004 study (see Fig. 8).
We note in passing that a Gaussian fit to the LF of the B2004 RR Lyraes yields
a 1-$\sigma$ width of 0.12 mag, which is comparable to the value for the
RR Lyraes in the present study.

\subsection{Number Ratios}

Of the 681 total RR Lyrae stars in our sample, 555 (267 in Field 1 and 288 in Field 2)
are of the ab-type with the remainder being c-type (57 in Field 1 and 69 in Field 2). 
This leads to a ratio of $N_{c}/N_{abc}$=0.19$\pm$0.02, which is quite different than
the value of $N_{c}/N_{abc}$=0.46$\pm$0.11 inferred from the B2004 data.
Therefore, our samples of ab-type and c-type RR Lyraes 
are a factor of $\sim$2 greater and a factor of $\sim$2 less, respectively, than 
what we would
expect based on the B2004 field. Taken at face value, this suggests 
that the old population in the environs of M31 exhibits different pulsation properties at 
4--6 kpc as compared with 11 kpc.

To evaluate the validity of this assertion, we need to address the question of
incompleteness in our sample of RR Lyraes. Are there significant numbers of
RR Lyraes in our fields that we have failed to identify?
We begin by noting that B2004 claim that their samples of 
c-type and ab-type RR Lyraes are complete and not significantly contaminated 
by dwarf cepheids. Therefore, we can gain some insight by comparing 
the amplitude distributions of the B2004 RR Lyrae variables
with our sample as shown in Fig. 9. 
This comparison suggests that the 
amplitude distribution of the B2004 RR Lyraes are consistent with those
of our sample. Application of the Kolmogorov-Smirnov test to these 
distributions confirms this suggestion.
The fact that, at the low amplitude end, the two distributions
are not substantially different argues that significant numbers of 
low amplitude RR Lyraes are not missing from our sample.



Another possibility to explain the differences in the $N_{c}/N_{abc}$ ratio
between our fields and the one at 11 kpc is that our Field 1 data are 
significantly influenced by the stellar populations
of M32 and not M31. In fact, we find $N_{c}/N_{abc}$=0.18$\pm$0.0.025 in Field 1 and
$N_{c}/N_{abc}$=0.19$\pm$0.025 in Field 2, which are statistically indistinguishable 
from each other. In addition, the density of RR Lyrae variables and their
period distributions are indistinguishable between fields 1 and 2. For example,
for the ab-type variables, Field 1 exhibits a mean period of
$\langle$$P_{ab}$$\rangle$$=0.553 \pm 0.004$ while Field 2 shows
$\langle$$P_{ab}$$\rangle$$=0.561 \pm 0.005$. In the case of the c-type
RR Lyraes, the analogous values are
$\langle$$P_{c}$$\rangle$$=0.326 \pm 0.005$ and
$\langle$$P_{c}$$\rangle$$=0.327 \pm 0.004$, respectively. These values
suggests that both of our fields sample the regions around
M31 and are minimally contaminated by M32 RR Lyraes.

If this difference in the $N_{c}/N_{abc}$ ratio between a radial distance of
11 kpc and 4--6 kpc in M31 is real, then it suggests that the RR Lyraes at
11 kpc are more akin to their brethren in Oosterhoff type II globular clusters while
those at 4--6 kpc are more like RR Lyraes in Oosterhoff type I clusters. This
is based on the fundings of Castellani et al. (2003) who examined the
$N_{c}/N_{abc}$ ratio in Galactic globulars. They found that among the 
12 clusters with 40 or more variables, 
$\langle$$N_{c}/N_{abc}$$\rangle$ = 0.37 for Oosterhoff II clusters and 
$\langle$$N_{c}/N_{abc}$$\rangle$ = 0.17 for those of Oosterhoff type I.
These compare favorably with the values of 
$\langle$$N_{c}/N_{abc}$$\rangle$ = 0.46 for the 11 kpc field and
$\langle$$N_{c}/N_{abc}$$\rangle$ = 0.19 for the 4--6 kpc field. 


\subsection{Periods, Amplitudes, and Metallicities}

The Bailey Diagram for our sample of RR Lyraes is shown in Fig. 10 where
we plot the variables in the two fields using different colors. However, it
is clear that they occupy the same regions of this diagram. The ab-type
RR Lyrae variables are shown with open circles while the c-type stars are
plotted as open triangles. The dashed
line in Fig. 10 represents the relation exhibited by the RRab stars in the
B2004 field. The solid lines are the relations for Oosterhoff I and II globular
clusters from Clement (2000). These lines have been adjusted to account
for the difference between an amplitude in the V-band and one in the F606W
band. Interestingly, the B2004 RR Lyrae relation
is closer to the OoI line even though the $N_{c}/N_{abc}$
ratio in the B2004 field is closer to that of OoII clusters. It is unclear why
this should be the case.

The B2004 line appears to be offset 
compared with our data suggesting slightly shorter  periods for the RRab
variables in our fields. This behavior is further exemplified in Fig. 11 which
shows the period distributions in the two fields compared with the mean
periods of the ab- and c-type RR Lyraes from B2004. 
We find mean periods of $\langle$$P_{ab}$$\rangle$$=0.557\pm0.003$ and
$\langle$$P_{c}$$\rangle$$=0.327\pm0.003$. For the B2004
field, these values are $\langle$$P_{ab}$$\rangle$$=0.594\pm0.015$ and
$\langle$$P_{c}$$\rangle$$=0.316\pm0.007$. The mean periods of the c-type variables 
are statistically indistinguishable from each other but the ab-types in our fields exhibit
a somewhat shorter period as compared with those in the B2004 fields.



It is well known that as the metallicity of ab-type RR Lyraes increases, their periods
decrease (e.g. Sandage 1993; Layden 1995; Sarajedini et al. 2006). Thus, the period 
distribution of these stars (Fig. 11) can be converted to a metallicity
distribution using equations derived by previous investigators. 
Using the data of Layden (2005, private communication) for 132 Galactic RR Lyraes
in the solar neighborhood, Sarajedini et al. (2006) established a relation between period and 
metal abundance of the form

\begin{eqnarray}
[Fe/H] = -3.43 - 7.82~Log~P_{ab}.
\end{eqnarray}

This equation does not take into account the amplitudes of the RR Lyraes 
even though, as Fig. 10 shows, there is a relation between amplitude and 
period for the ab-types. The work
of Alcock et al. (2000) yielded a period-amplitude-metallicity relation of the form

\begin{eqnarray}
[Fe/H] = -8.85[Log~P_{ab} + 0.15A(V)] -2.60,
\end{eqnarray}

\noindent where $A(V)$ represents the amplitude in the V-band. We applied an 8\% increase to the
$m_{F606W}$ amplitudes to convert them to V-band values (B2004).
Figure 12 shows the metallicity
distribution function (MDF) for the ab-type RR Lyraes. The top panel compares the results
obtained using equations (1) and (2) while the lower panel compares the MDFs for
the two fields using Equation (2). 

We see in the upper panel of Fig. 12 that, while the two MDFs exhibit very similar
peak metallicities, the MDF generated using Eqn (2) displays a more prominent peak. This is
representative of the fact that Eqn (2) accounts for the variation of period with amplitude
as well as metallicity resulting in a cleaner abundance signature in the MDF. In the
lower panel of Fig. 12, we see that the peaks of the RR Lyrae MDFs in our two fields differ
by $\sim$0.1 dex; we find $\langle$[Fe/H]$\rangle$ = $-1.46\pm0.03$ for Field 1 and 
$\langle$[Fe/H]$\rangle$ = $-1.54\pm0.03$ for Field 2, where the errors represent standard
errors of the mean. This difference is not statistically significant.  

Combining the RR Lyraes in the two fields yields the MDF shown as the solid line
in Fig. 13, wherein the binned and generalized histograms are shown. The latter has
been constructed using an error of 0.31 dex per star (Alcock et al. 2000). The peak
metallicity for our sample of ab-type RR Lyraes is then 
$\langle$[Fe/H]$\rangle$=$-1.50\pm0.02$ 
where the error is the standard error of the mean. The systematic error of this
measurement is likely to be closer to $\sim$0.3 dex. The dotted distributions in
Fig. 13 are the binned and generalized histograms for the RRab stars in the sample of
B2004 scaled to the same number of ab-type RR Lyraes as in our fields. The peak abundance 
of this MDF is $\langle$[Fe/H]$\rangle$=$-1.77\pm0.06$. 

There are three observations we can make regarding the appearance of Fig. 13. First,
it would seem that the errors on the individual metallicity measurements are
significant enough to overwhelm any fine-structure that may be present in the
two MDFs. That is to say, both MDFs look essentially like normal distributions.
Second, since we have applied the same transformation from period to metallicity 
to both sets of RR Lyraes, we can assert with significant statistical certainty that
the mean metal abundance of the RR Lyraes in the B2004 field is lower than
that of the RR Lyraes in our two fields. This difference amounts to 
$\Delta$$[Fe/H] = 0.27 \pm 0.06$ and reflects back on the period
shift seen in the Bailey Diagram shown in Fig. 10. Third, there are a small
but non-negligible number of ab-type RR Lyraes with metallicities above
$[Fe/H]$$\sim$-1 that are not seen in the B2004 field. Given the fact that
our field is closer to the central regions of M31 as compared with the B2004, it
is possible that these metal-rich RR Lyraes could belong to the bulge or
disk of M31. We return to this point in the next section.

One last point needs to be addressed before leaving this section and that is
concerned with the M31 distance implied by the RR Lyraes in our sample. Using
the relation advocated by Chaboyer (1999) of $M_V(RR) = 0.23[Fe/H] + 0.93$
and a reddening of $E(B-V) = 0.08 \pm 0.03$ (Schlegel et al. 1998), 
we find a distance modulus of $(m-M)_0=24.46\pm0.11$. This is consistent
with the B2004 value and a number of previous determinations.

\section{Discussion}

We now seek to place our RR Lyrae abundance results within the broader context of
the projected radial metallicity distribution of various populations in the environs
of M31. Figure 14 shows this information for a range of stellar populations in and
around M31. The metallicities for the RR Lyraes in the two fields considered herein
are shown by the filled circles while the open circle represents the RR Lyraes
in the B2004 field. The inner-most point is the bulge metallicity from the work of
Sarajedini \& Jablonka (2005), while the remaining open squares are the bulge/halo
points from the work of Kalirai et al. (2006) as shown in their Table 3. The dashed
line is the least squares fit to the open squares with a slope of $-0.75 \pm 011$. 
The other points represent
the dwarf spheroidal companions to M31 (crosses, Grebel et al. 2003; 
Koch \& Grebel 2006), the globular cluster G1 (filled square, Meylan et al. 2001),
and the furthest globular cluster in M31 (open triangle, Martin et al. 2006). Note
that we have adopted the mean metallicity of M32 from the work of Grillmair
et al. (1996). The
elongated rectangle represents the locations of the halo globular clusters
in M33 from Sarajedini et al. (2000). All of these values are based on a 
distance of $(m-M)_o = 24.43$ (770 kpc) for M31. In
cases where an error in the metallicity is not available, we have adopted a value
of $\pm$0.2 dex.

We see in Fig. 14 a clear representation of the notion that the halo population
in M31 does not begin to dominate until a galactocentric distance of $\sim$30 kpc,
as suggested by a number of authors (Guhathakurta et al. 2005; Irwin et al. 2005;
Kalirai et al. 2006; Koch et al. 2007). At this location, we see
a transition region between the globular cluster G1 which is consistent with the 
inner-spheroid metallicity gradient (dashed line) and the dwarf spheroidal
galaxies which show no relation between abundance and galactocentric
distance. 
In this sense, it would appear that the RR Lyrae populations in our
field and that of B2004 follow the trend outlined by the stellar populations
outside of $\sim$30 kpc. This suggests that the RR Lyraes at these locations
are probably members of the M31 halo rather than its bulge suggesting that
the halo can be studied as close as 4 kpc from the center of M31 by
focusing on the RR Lyraes.

\section{Summary and Conclusions}

We have presented F606W ($\sim$V) observations from the HST archive taken 
with the Advanced Camera for Surveys of two fields located 4-6 kpc from
the center of  M31. 
In these regions, we identify 752 variable stars of which 681 are likely to be
{\it bona fide} RR Lyraes. From the properties of these stars, we draw the 
following conclusions.

\noindent 1) The mean magnitude of the RR Lyrae stars is 
$\langle$$V$$\rangle$$=25.29 \pm 0.05$ where the uncertainty combines both
the random and systematic errors. This is in good agreement with the
results of Brown et al. (2004) 

\noindent 2) The ratio of c-type RR Lyraes to all types is reminiscent of 
the RR Lyraes in Oosterhoff type I globular clusters in the Milky Way. This
ratio is significantly different than the Brown et al. (2004) field at 11 kpc from
the center of M31 wherein this ratio is closer to that of Oosterhoff II clusters.

\noindent 3) When the periods and amplitudes of the ab-type RR Lyraes in our
sample are interpreted in terms of metallicity, we find the metallicity
distribution function to be indistinguishable from a Gaussian with a 
peak at $\langle$[Fe/H]$\rangle$=$-1.50\pm0.02$,
where the error is the standard error of the mean. The same analysis
applied to the Brown et al. (2004) RR Lyraes yields a peak of
$\langle$[Fe/H]$\rangle$=$-1.77\pm0.06$. 

\noindent 4) Using the RR Lyrae luminosity - metallicity
relation advocated by Chaboyer (1999) and a reddening of $E(B-V) = 0.08 \pm 0.03$, 
we find a distance modulus of $(m-M)_0=24.46\pm0.11$ for M31. 

\noindent 5) We examine the radial metallicity gradient in the
environs of M31 using published values for the bulge and halo of M31
as well as the abundances of the dwarf spheroidal companions and globular
clusters of M31. In this context, despite the relative proximity of the RR Lyraes
in the present study to the center of M31, their metal abundance is 
more reminiscent of a halo population than a bulge or disk. Therefore,
by using the RR Lyraes as a proxy, the halo can be studied as 
close as 4 kpc from the center of M31.

\acknowledgements
We are grateful to Andy Layden, Nathan De Lee, and Karen Kinemuchi for useful
conversations as this manuscript was being written. A. S.
is grateful for support from NASA through grant AR-11277.01-A from the Space 
Telescope Science Institute, which is operated by the Association of Universities
for Research in Astronomy, Inc., for NASA under contract NAS5-26555.

\clearpage

\begin{deluxetable}{clccccccc}
\tablecaption{Observing Log}
\tabletypesize{\scriptsize}
\tablewidth{0pt}
\tablehead{
\colhead{Field}
  &\colhead{RA (J2000)}
  &\colhead{Decl. (J2000)}
  &\colhead{Starting Date}
  &\colhead{Datasets}
  &\colhead{Filter}
  &\colhead{Exp Time}
  &\colhead{HJD Range (+2 453 000)}
}
\startdata
1 & 00 42 41.2 & 40 46 38 & September 22, 2005 & J9H905, J9H906, & F606W & 16 $\times$ 1136s, & 635.97 to 638.25\\
  &                    &                &                                   & J9H907, J9H908 &             & 16 $\times$ 1177s &  \\
2 & 00 43 20.8 & 40 57 25 & February 9, 2006 & J9H913, J9H914,  & F606W & 16 $\times$ 1136s, &  775.88 to 778.96 \\
  &                    &                &                             &  J9H915, J9H916 &               &  16 $\times$ 1177s&   \\
\enddata
\end{deluxetable}

\input{alldata_field1_page1.tex}
\input{alldata_field2_page1.tex}

\begin{deluxetable}{ccccccl}
\tablecaption{Field 1 Variable Stars}
\tabletypesize{\footnotesize}
\tablewidth{0pt}
\tablehead{
\colhead{Star ID}
   &\colhead{RA (J2000)}
   &\colhead{Dec (J2000)}
   &\colhead{Period (days)}
   &\colhead{$m_{F606W}$ Amplitude}
   &\colhead{$\langle$$m_{F606W}$$\rangle$}
   &\colhead{Type}
 }
\startdata
     1 &  0 42 38.35 & 40 47  8.3 & 1.551 & 0.301 & 21.337 & long period \\ 
     2 &  0 42 41.92 & 40 45 19.1 & 2.559 & 0.919 & 22.070 & long period \\ 
     3 &  0 42 37.94 & 40 45 18.9 & 1.408 & 0.100 & 24.029 & long period \\ 
     4 &  0 42 34.90 & 40 46  4.3 & 0.387 & 0.288 & 24.712 & RRc         \\ 
     5 &  0 42 42.82 & 40 45 23.6 & 1.561 & 0.490 & 24.685 & Eclipsing   \\ 
     6 &  0 42 34.49 & 40 46 29.0 & 1.400 & 0.246 & 24.758 & Contact     \\ 
     7 &  0 42 41.61 & 40 46  7.4 & 0.660 & 0.725 & 24.877 & RRab        \\ 
     8 &  0 42 42.42 & 40 44 59.8 & 1.539 & 0.160 & 24.820 & long period \\
\enddata
\end{deluxetable}

\begin{deluxetable}{ccccccl}
\tablecaption{Field 2 Variable Stars}
\tabletypesize{\footnotesize}
\tablewidth{0pt}
\tablehead{
\colhead{Star ID}
   &\colhead{RA (J2000)}
   &\colhead{Dec (J2000)}
   &\colhead{Period (days)}
   &\colhead{$m_{F606W}$ Amplitude}
   &\colhead{$\langle$$m_{F606W}$$\rangle$}
   &\colhead{Type}
 }
\startdata
   366 &  0 43 17.55 & 40 58 42.3 & 1.886 & 0.314 & 21.732 & long period        \\ 
   367 &  0 43 20.31 & 40 58 12.1 & 2.234 & 0.387 & 22.141 & long period        \\ 
   368 &  0 43 24.43 & 40 56 27.8 & 2.220 & 0.288 & 22.146 & long period        \\ 
   369 &  0 43 24.57 & 40 56 51.9 & 1.366 & 0.501 & 23.625 & long period        \\ 
   370 &  0 43 16.06 & 40 58 29.2 & 1.978 & 0.291 & 23.675 & Eclipsing          \\ 
   371 &  0 43 23.53 & 40 57  3.9 & 0.557 & 0.558 & 23.794 & RRab               \\ 
   372 &  0 43 30.90 & 40 56 53.4 & 2.575 & 0.794 & 24.014 & long period        \\ 
   373 &  0 43 17.54 & 40 58 48.8 & 1.142 & 0.839 & 24.299 & Eclipsing          \\
\enddata
\end{deluxetable}



\clearpage
\begin{figure}
\epsscale{0.8}
\caption{The location of our ACS fields overplotted on
a digitized sky survey image in the region of M31. The dwarf elliptical galaxy M32 is near
the center of the image. North is up and east is to the left.}
\end{figure}

\clearpage
\begin{figure}
\epsscale{0.8}
\plotone{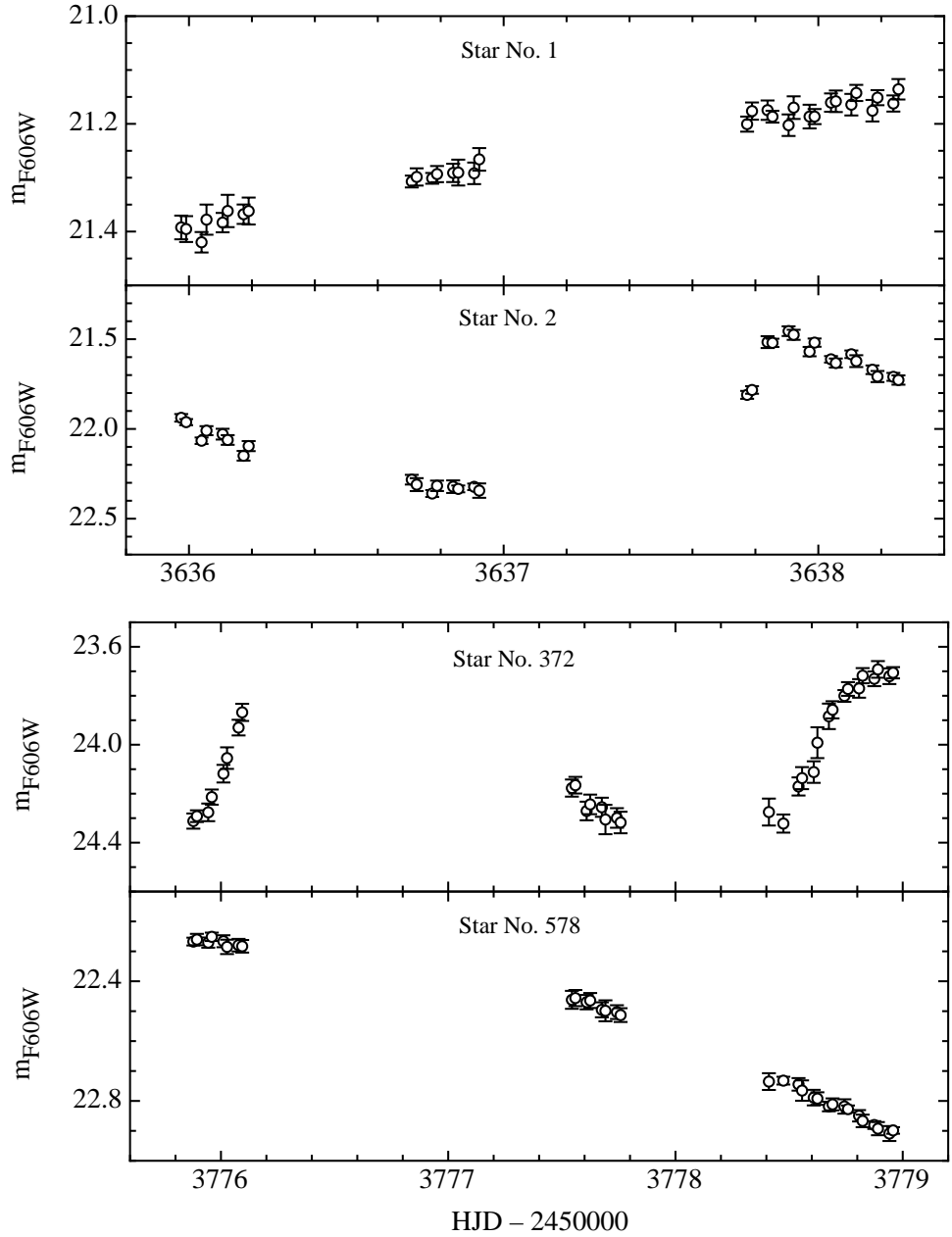}
\caption{Unphased light curves for a sample of our variable stars with periods that are 
comparable to or longer than our observing window. The numbers refer to the stars in 
Tables 4 and 5.}
\end{figure}

\clearpage
\begin{figure}
\epsscale{0.8}
\plotone{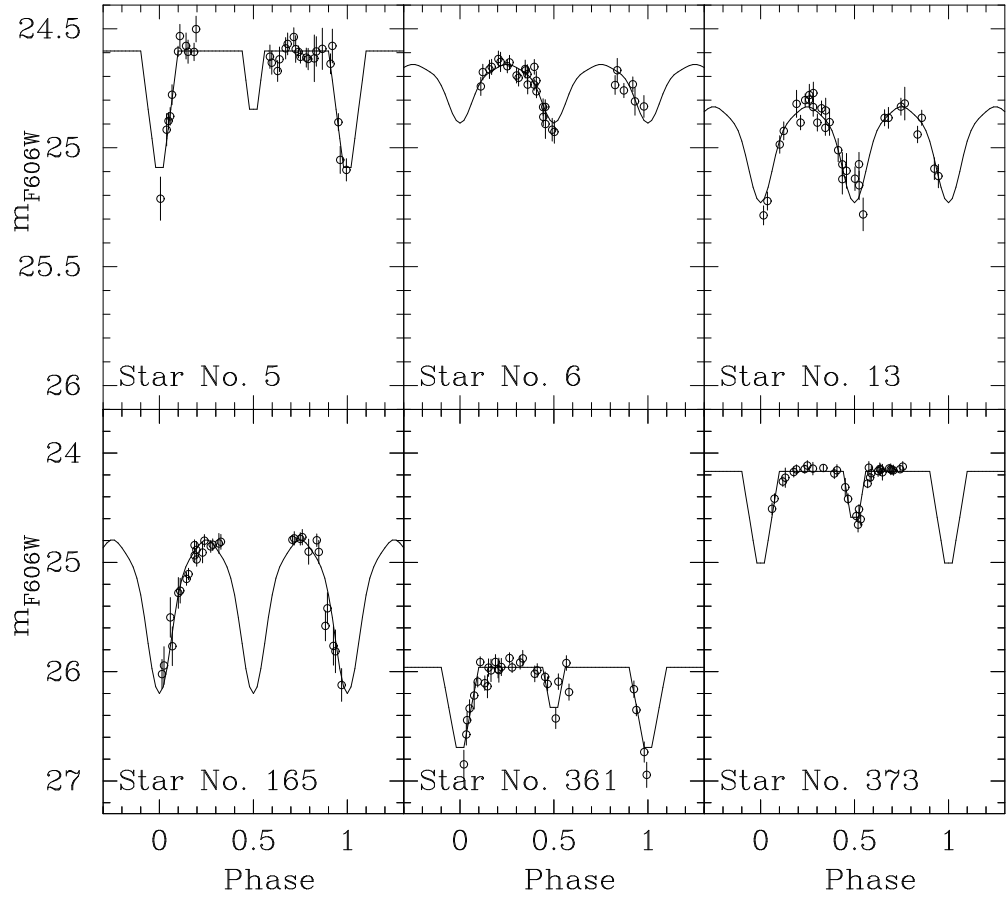}
\caption{Phased light curves for some of the contact and eclipsing binaries identified in this study.}
\end{figure}

\clearpage
\begin{figure}
\epsscale{0.85}
\plotone{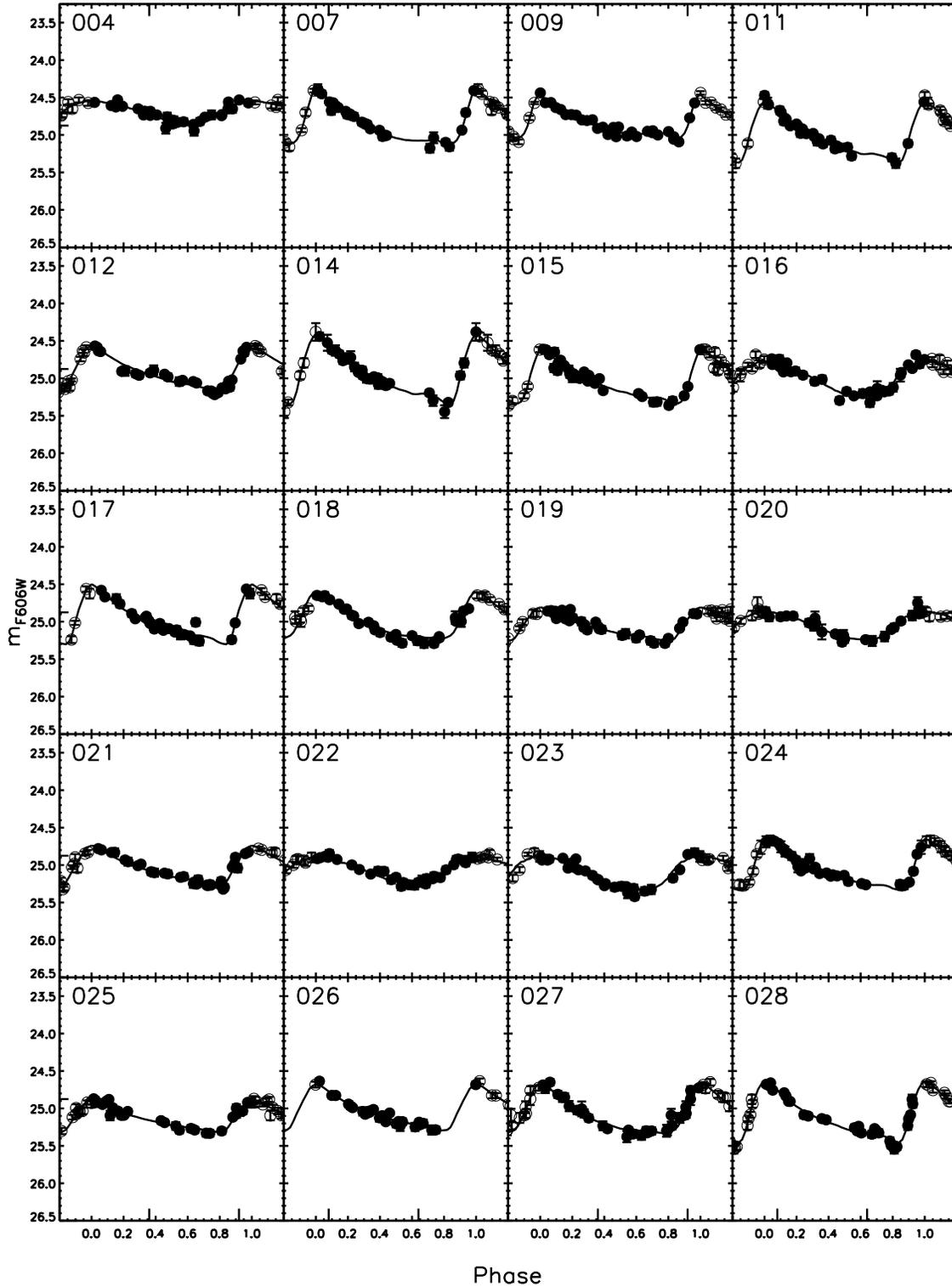}
\caption{Phased light curves for the 681 RR Lyrae variables identified in this study. The
open circles are repeated to complete the light curve for phase less than zero and greater
than one.}
\end{figure}



\clearpage
\begin{figure}
\epsscale{0.8}
\plotone{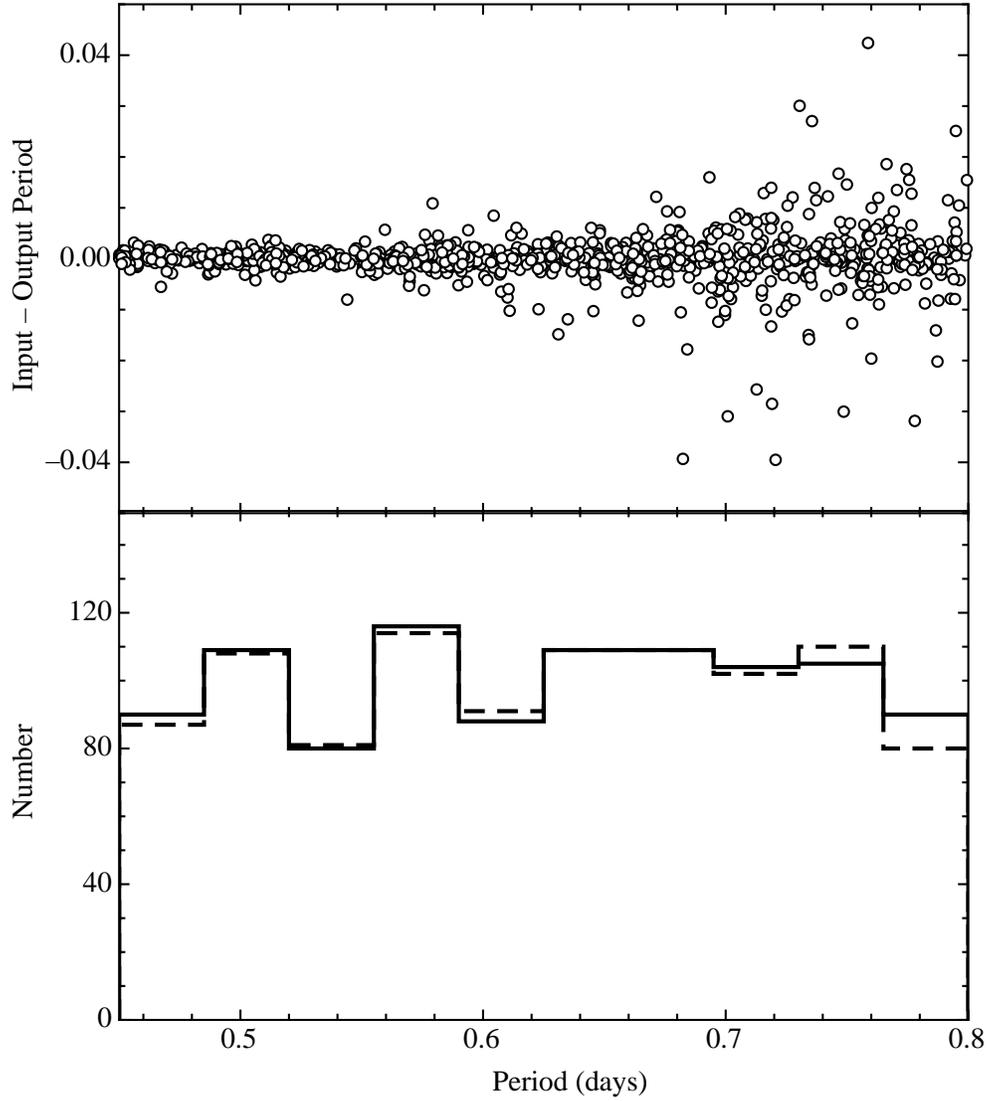}
\caption{The results of the light curve simulations performed in order to characterize
any biases in our period
finding algorithm in the case of the RRab variables. The upper panel plots the
variation of input minus recovered period while the lower panel plots the period distribution
both as a function of period in days. The solid line is the input distribution while the
dashed line represents the recovered one. These simulations suggest that the combination
of the input data and the period-finding algorithm do not introduce significant biases in
our derived periods. }
\end{figure}

\clearpage
\begin{figure}
\epsscale{0.8}
\plotone{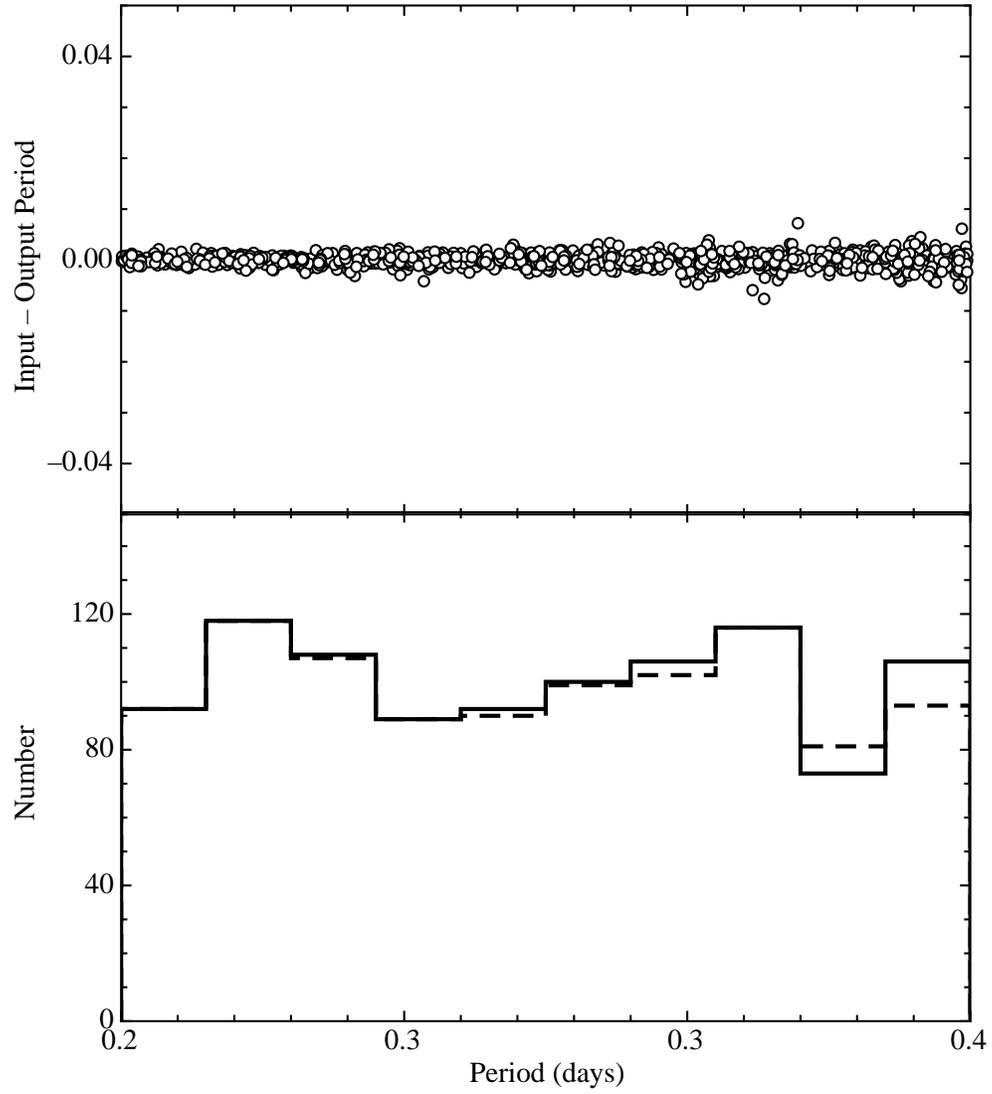}
\caption{Same as Fig. 6 except that the simulations have been performed using
an c-type RR Lyrae variable light curve.}
\end{figure}

\clearpage
\begin{figure}
\epsscale{0.75}
\caption{The luminosity functions (LFs) for the non-variable stars in the two fields studied herein. The
solid line in the lower panel is the F606W band LF for the field observed in 2005 (field 1 in Table 1) 
while the dashed line is the LF for the region observed in 2006 (field 2 in Table 1). The two upper
panels show the variation of the photometric error as output by ALLFRAME with magnitude. We have plotted every 10th point to make the appearance of
these plots manageable.}
\end{figure}

\clearpage
\begin{figure}
\epsscale{0.8}
\plotone{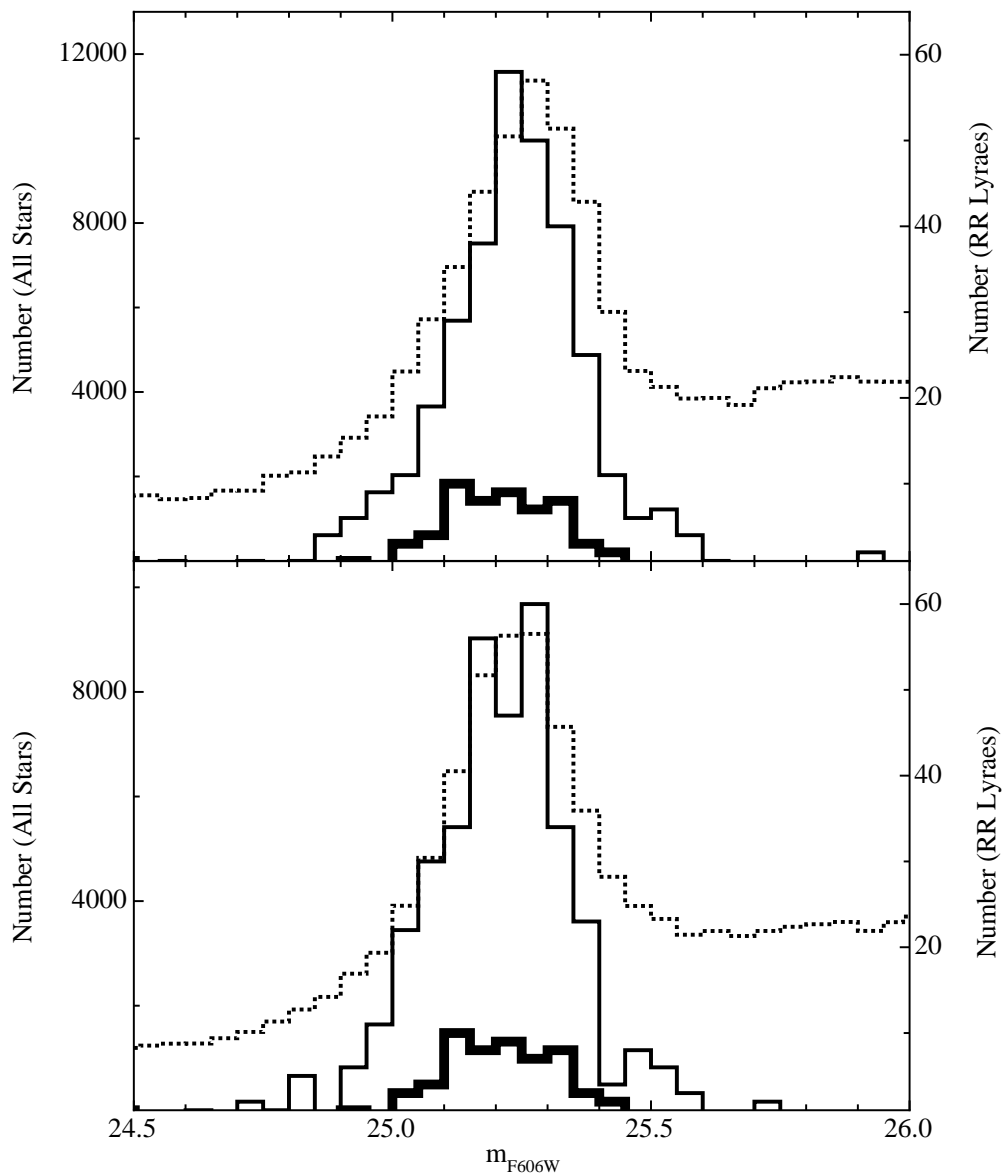}
\caption{The luminosity functions (LFs) for the non-variable stars (dashed lines) compared with 
those of the RR Lyrae candidates (thin solid lines) in the two fields studied herein (see Table 1). 
The thicker solid lines represent the RR Lyrae stars from the study of Brown et al. (2004).}
\end{figure}

\clearpage
\begin{figure}
\epsscale{0.8}
\plotone{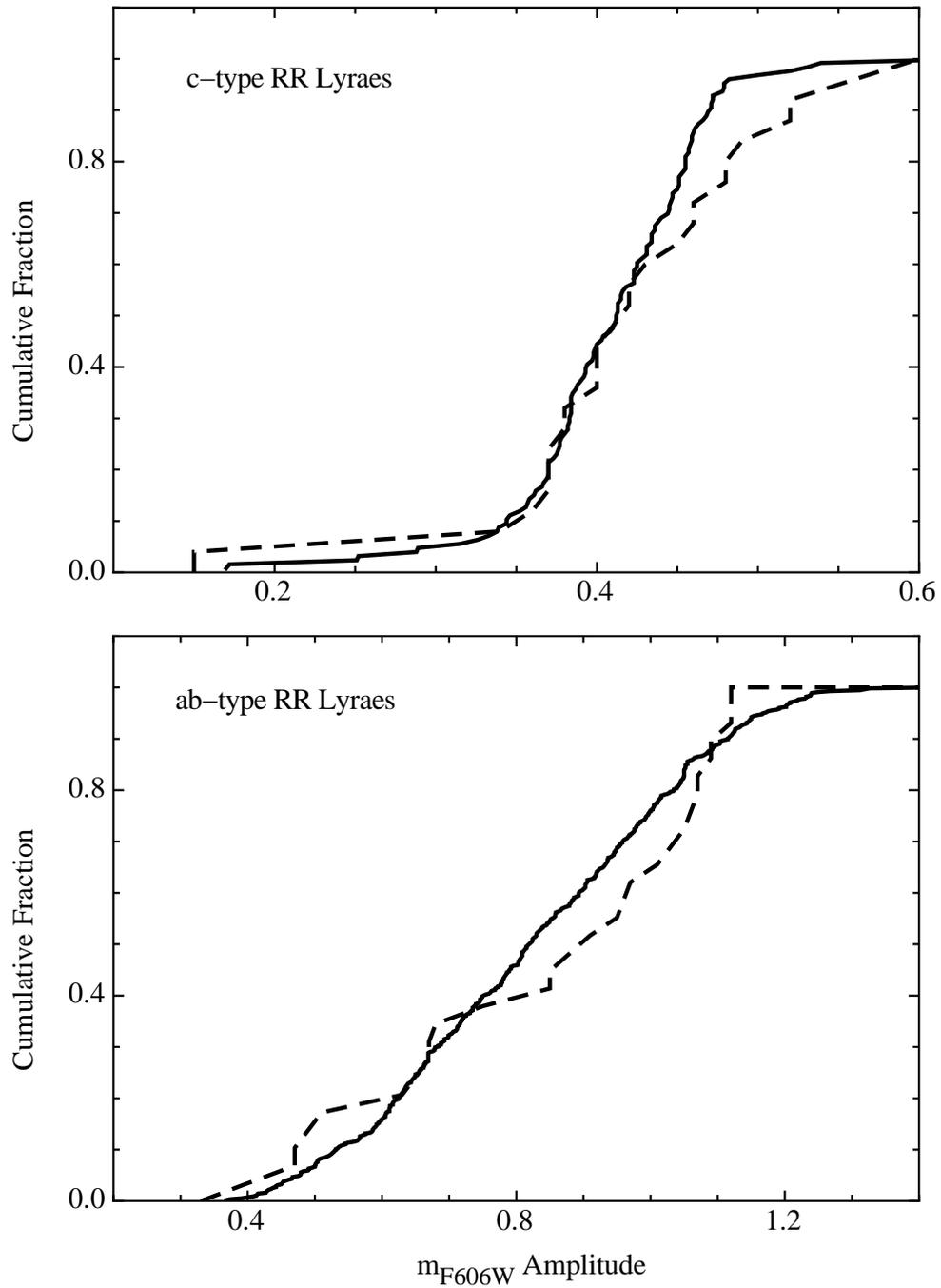}
\caption{The upper panel shows the amplitude distribution of the c-type RR Lyraes
from the present study (solid line) and the work of 
Brown et al. (2004, dotted line) scaled to the same maximum.
The lower panel is the same as the upper one except that the ab-type RR Lyraes
are plotted.}
\end{figure}

\clearpage
\begin{figure}
\epsscale{0.8}
\plotone{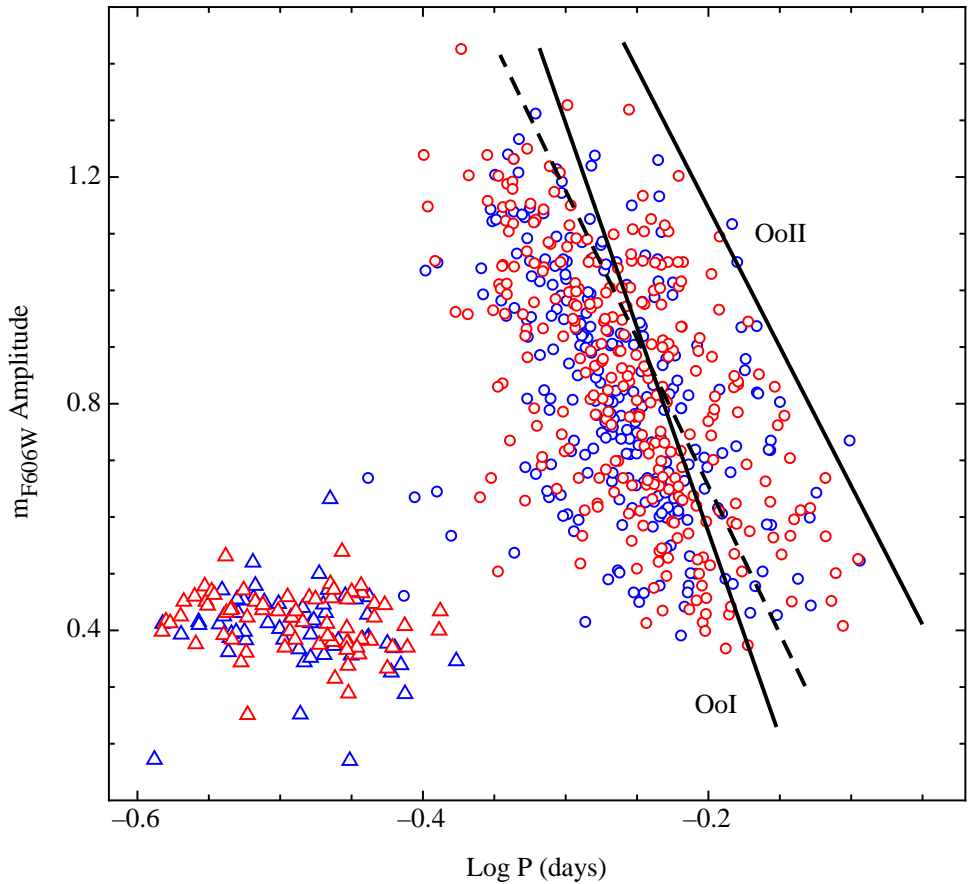}
\caption{The Bailey Diagram for the RR Lyrae candidates in our fields. The open circles are
the ab-type RR Lyraes while the open triangles represent the c-type variables. The color coding
represents the field number with blue being stars observed in field 1 and red those found
in field 2. This plot shows that, while the ab- and c-type RR Lyraes occupy their characteristic
locations in this diagram, there is no significant difference between the RR Lyraes in
the two observed fields. The dashed line shows the relation exhibited by the RRab stars
in the field observed by Brown et al. (2004). The solid lines are the relations for Oosterhoff
type I and II globular clusters from Clement (2000). These lines have been adjusted to account
for the difference between an amplitude in the V-band and one in the F606W
band.}
\end{figure}

\clearpage
\begin{figure}
\epsscale{0.8}
\plotone{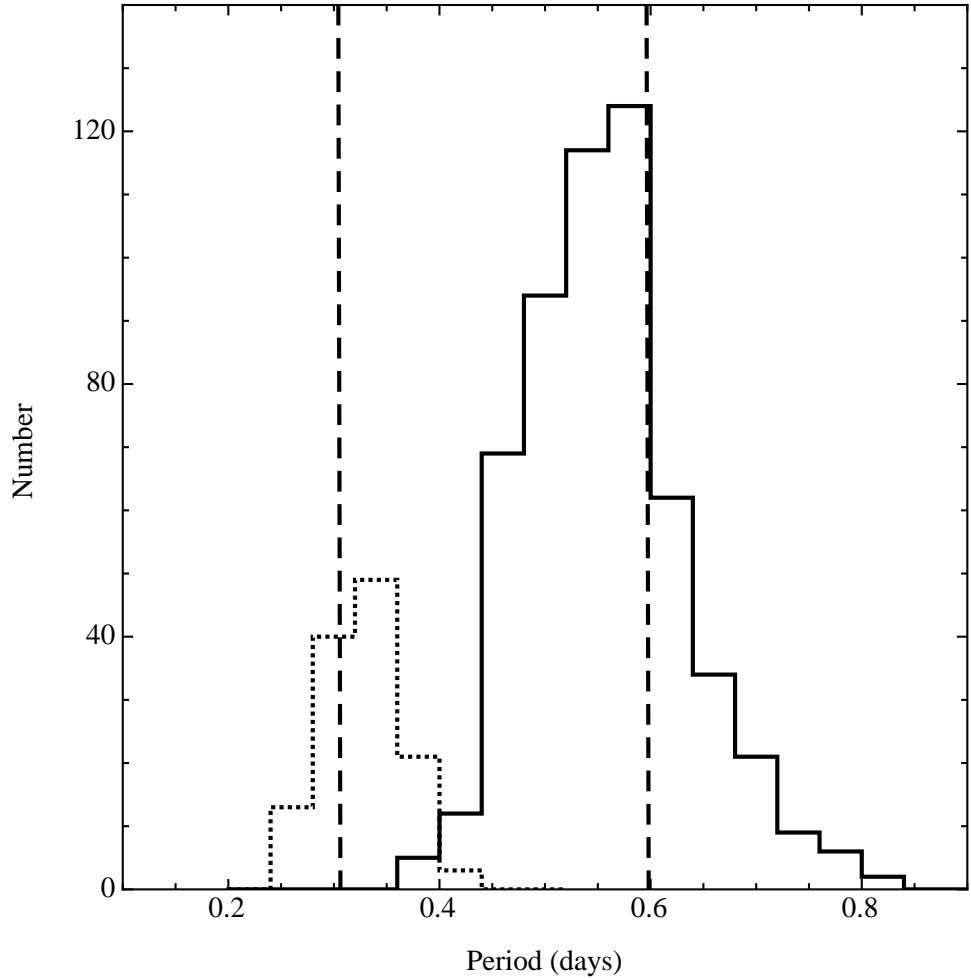}
\caption{The period distributions of the ab-type (solid) and c-type (dotted) RR Lyrae variables.
The dashed vertical lines represent the mean periods of these same stars from Brown et al.
(2004). While the mean periods of the c-type variables agree between our study and that of
Brown et al. (2004), the mean period of the ab-type RR Lyraes is somewhat shorter in our
sample as compared with that of Brown et al.}
\end{figure}

\clearpage
\begin{figure}
\epsscale{0.8}
\plotone{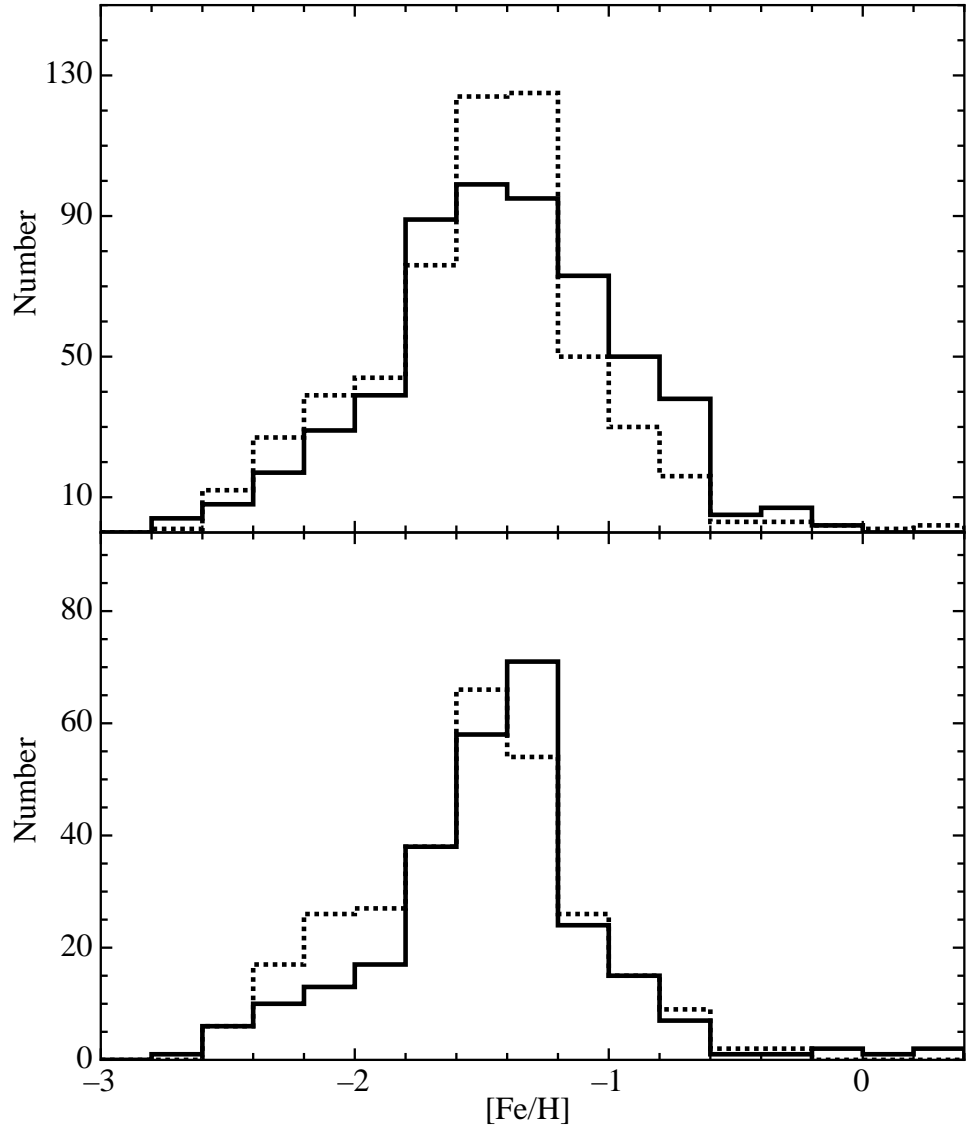}
\caption{The upper panel shows the metallicity distribution function for our sample of RRab
variables using two different formulations for the conversion between period and metal 
abundance (solid line - Eqn (1), dashed line - Eqn (2)). The lower panel shows the MDFs
derived using Eqn (2) for the two observed fields (dashed line - Field 2, solid line - Field 1).}
\end{figure}

\clearpage
\begin{figure}
\epsscale{0.9}
\plotone{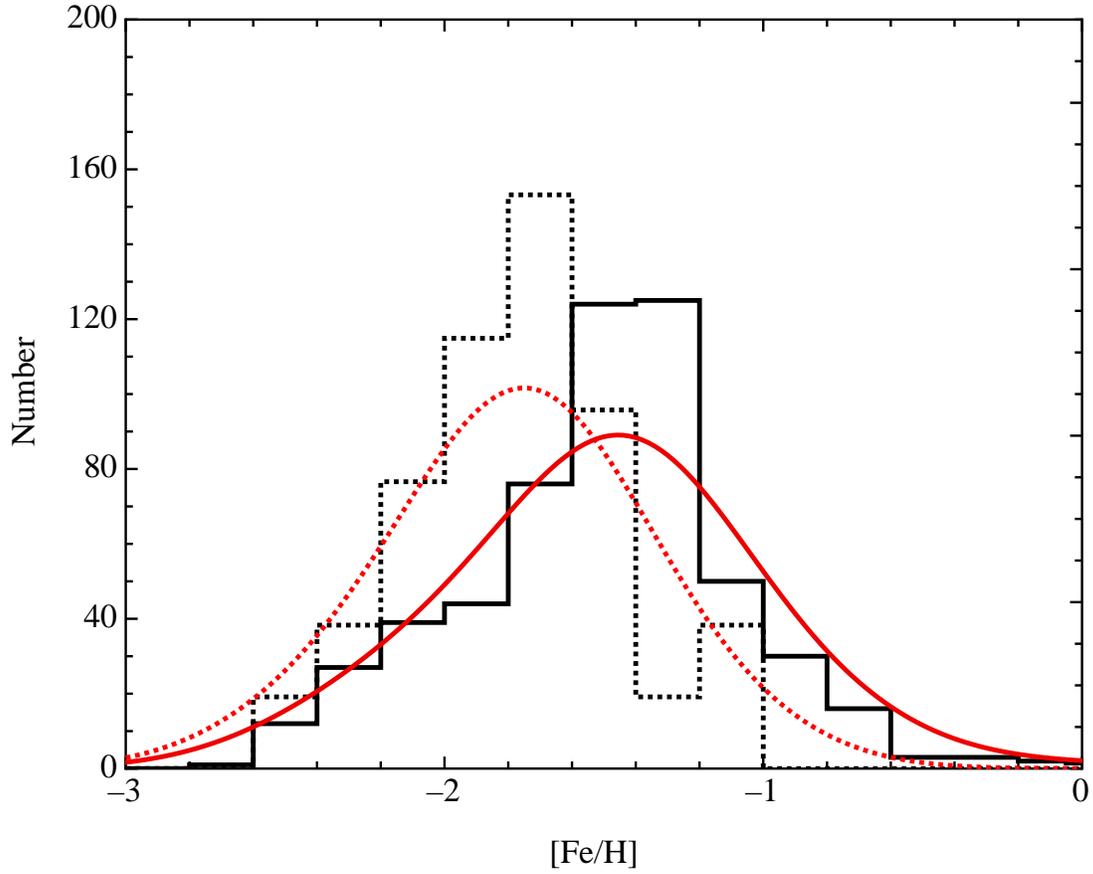}
\caption{A comparison of the metallicity distribution function derived from the ab-type
RR Lyraes from the present study (solid lines) and those from the Brown et al. (2004)
study (dotted lines). The latter has been scaled to match the number of ab-type
RR Lyraes in our two fields. Both binned and generalized histograms are shown.}
\end{figure}

\clearpage
\begin{figure}
\epsscale{0.9}
\plotone{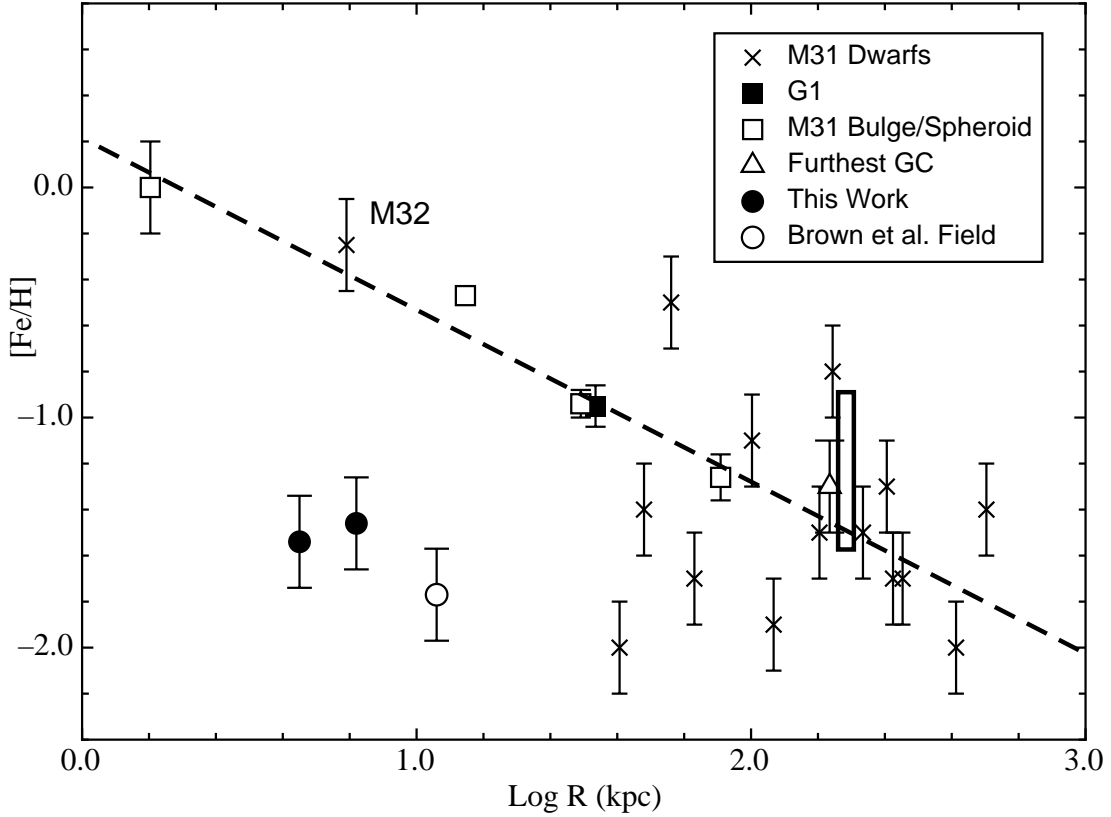}
\caption{A plot of the variation of metal abundance with projected distance from the center of M31. 
Our results are shown by the filled circles while the mean metallicity of the 
RR Lyraes studied by Brown et al. (2004) is indicated by the open circle. The
inner most open square represents the bulge abundance measured by Sarajedini \&
Jablonka (2006). The remaining open squares are the bulge/halo points from the work of
Kalirai et al. (2006). The dashed line is the least squares fit to these data with a slope
of --0.75 $\pm$ 0.11. The crosses
represent the dwarf galaxies surrounding M31 from the work of Grebel et al.
(2003) and Koch \& Grebel (2006) whereas the abundance of M32 is taken from
Grillmair et al. (1996). The filled square is the well-known massive globular cluster
G1 studied by Meylan et al. (2001). The open triangle is the furthest known globular cluster
in M31 discovered by Martin et al. (2006). 
For completeness, the boxed region shows the location of the halo globular clusters in M33
from the work of Sarajedini et al. (2000). All of these points have been scaled to an
M31 distance of $(m-M)_0$ = 24.43.}
\end{figure}

\end{document}

%% file: alldata_field1_page1.tex
 \begin{deluxetable}{ccccccccccc}
 \tablecaption{Raw Light Curves Field 1}
 \tabletypesize{\tiny}
 \tablewidth{0pt}
 \tablehead{
 \colhead{Epoch}
 & \colhead{001 Mag} & \colhead{001 Err}
 & \colhead{002 Mag} & \colhead{002 Err}
 & \colhead{003 Mag} & \colhead{003 Err}
 & \colhead{004 Mag} & \colhead{004 Err}
 & \colhead{005 Mag} & \colhead{005 Err}
 }
 \startdata
3635.97534180 &  21.414 &   0.022 &  21.960 &   0.021 &  24.033 &   0.033 &  24.782 &   0.059 &  24.639 &   0.049 \\
3635.99096680 &  21.417 &   0.024 &  21.984 &   0.017 &  24.023 &   0.033 &  24.826 &   0.052 &  24.664 &   0.049 \\
3636.04003906 &  21.442 &   0.019 &  22.088 &   0.018 &  24.024 &   0.052 &  24.967 &   0.065 &  24.699 &   0.044 \\
3636.05566406 &  21.400 &   0.028 &  22.032 &   0.025 &  24.011 &   0.021 & \nodata & \nodata &  24.650 &   0.052 \\
3636.10693359 &  21.405 &   0.018 &  22.051 &   0.030 &  24.016 &   0.051 &  24.750 &   0.034 &  24.604 &   0.045 \\
3636.12304688 &  21.384 &   0.030 &  22.083 &   0.027 &  24.013 &   0.042 &  24.582 &   0.038 &  24.585 &   0.046 \\
3636.17358398 &  21.390 &   0.018 &  22.171 &   0.027 &  23.998 &   0.043 & \nodata & \nodata &  24.556 &   0.039 \\
3636.18969727 &  21.384 &   0.025 &  22.118 &   0.028 & \nodata & \nodata & \nodata & \nodata &  24.607 &   0.056 \\
\enddata
 \end{deluxetable}

%% file: alldata_field2_page1.tex
\begin{deluxetable}{ccccccccccc}
 \tablecaption{Raw Light Curves Field 2}
 \tabletypesize{\tiny}
 \tablewidth{0pt}
 \tablehead{
 \colhead{Epoch}
 & \colhead{366 Mag} & \colhead{366 Err}
 & \colhead{367 Mag} & \colhead{367 Err}
 & \colhead{368 Mag} & \colhead{368 Err}
 & \colhead{369 Mag} & \colhead{369 Err}
 & \colhead{370 Mag} & \colhead{370 Err}
 }
 \startdata
3775.87866211 &  21.896 &   0.034 &  22.464 &   0.029 &  22.360 &   0.038 & \nodata & \nodata &  23.676 &   0.023 \\
3775.89428711 &  21.955 &   0.031 &  22.409 &   0.050 &  22.387 &   0.038 &  23.583 &   0.025 &  23.667 &   0.036 \\
3775.94409180 &  21.901 &   0.024 &  22.388 &   0.041 &  22.347 &   0.028 &  23.624 &   0.024 &  23.597 &   0.026 \\
3775.95971680 &  21.885 &   0.035 &  22.443 &   0.029 &  22.342 &   0.038 &  23.626 &   0.026 &  23.664 &   0.030 \\
3776.01098633 &  21.930 &   0.028 &  22.417 &   0.043 &  22.347 &   0.029 &  23.686 &   0.027 &  23.607 &   0.020 \\
3776.02685547 &  21.881 &   0.034 &  22.400 &   0.032 &  22.387 &   0.032 &  23.689 &   0.036 &  23.579 &   0.036 \\
3776.07763672 &  21.886 &   0.027 &  22.384 &   0.032 &  22.391 &   0.035 &  23.716 &   0.026 &  23.584 &   0.033 \\
3776.09350586 &  21.914 &   0.019 &  22.323 &   0.063 &  22.332 &   0.028 &  23.688 &   0.037 &  23.592 &   0.022 \\
\enddata
 \end{deluxetable}